\documentclass[12pt]{article}
\usepackage{amssymb}
\newcommand{\bee}{\begin{eqnarray}}
\newcommand{\ede}{\end{eqnarray}}
\begin{document}
\baselineskip=12 pt
\begin{center}
{The Spin-weighted Spheroidal Wave functions in the Case of $s=1/2$
}
\end{center}
\begin{center}
Kun Dong ,${}$\footnote{e-mail : woailiuyanbin1@126.com}, Guihua
Tian ${}$\footnote{e-mail : hua2007@126.com},Yue
Sun${}$\footnote{e-mail : sunyue1101@126.com}
\end{center}
\begin{center}
School of Sciences,\\
Beijing University of Posts and Telecommunications,\\
Beijing, China, 100876.\\
\end{center}
\begin{abstract}
The spin-weighted spheroidal equations in the case $s=1/2$ is
thoroughly studied in the paper
  by means of the perturbation method in supersymmetry quantum mechanics.
   The first-five terms of the super-potential in the series of the parameter $\beta$ are
   given. The general form of the nth term of the superpotential is
   also obtained, which could derived from the previous terms
   $W_{k}$, $k<n$. From the results, it is easy to give the ground
   eigenfunction of the equation. Furthermore, the shape-invariance
   property is investigated in the series form of the parameter
   $\beta$ and is proven kept in this series form for the equations.
   This nice property guarantee one could obtain the excited
   eigenfunctions in the series form from the ground eigenfunctions
   by the method in supersymmetry quantum mechanics. This shows the
   perturbation method method in supersymmetry quantum mechanics
   could solve the spin-weight spheroidal wave equations completely in the
   series form of the small parameter $\beta$.
\end{abstract}

\section{Introduction}
The spin-weighted spheroidal functions first appeared in the study
of the stable problem of Kerr black hole.
  For the wave equation of the perturbation of $\varphi$ of Kerr black
  hole[1-2]
\begin{eqnarray}
&&\bigg[\frac{\left(r^2+a^2\right)}{\Delta}-a^2\sin^2\theta\bigg]\frac{\partial^2\psi
}{\partial t^2}+\frac{4Mar}{\Delta}\frac{\partial^2\psi }{\partial
t\partial
\phi}+\bigg[\frac{a^2}{\Delta}-\frac1{\sin^2\theta}\bigg]\frac{\partial^2\psi
}{\partial \phi^2}\nonumber\\
&& -\frac{1}{\sin \theta}\frac{\partial}{\partial\theta}\left(\sin
\theta \frac{\partial \psi }{\partial \theta}\right)
-2s\bigg[\frac{a\left(r-M\right)}{\Delta}+ \frac{i\cos
\theta}{\sin^2\theta}\bigg]\frac{\partial \psi }{\partial
\phi} \nonumber\\
&& - \Delta^{-s}\frac{\partial}{\partial
r}\left(\Delta^{s+1}\frac{\partial \psi }{\partial r}\right)
-2s\bigg[\frac{M\left(r^2-a^2\right)}{\Delta}-r-ia\cos\theta
\bigg]\frac{\partial \psi }{\partial
t}\nonumber\\
&& +\left(s^2\cot^2\theta-s\right)\psi =0 \label{Teukolsky}
\end{eqnarray}
Teukolsky  made nice separation
 for Eq.(\ref{Teukolsky}) and obtained radial equation for $R(r)$as
\begin{eqnarray}
&&\Delta^{-s}\frac{d}{d r}\left(\Delta^{s+1}\frac{dR }{d r}\right)\nonumber\\
&+&\bigg[\frac{K^2-2is\left(r-M\right)K}{\Delta}+ 4is\omega
r-\bar{\lambda} \bigg]R =0,\end{eqnarray} and the  angular equations
for $\Theta(\theta)$ as

\begin{eqnarray}&& \bigg[\frac{1}{\sin
\theta}\frac{d}{d\theta}\bigg(\sin \theta \frac{d}{d
\theta}\bigg)+s+a^{2}\omega ^2\cos ^2 \theta \nonumber \\ &-&
2as\omega \cos \theta -\frac{\left(m+s\cos \theta\right)^{2}}{\sin
^2 \theta}+E\bigg]\Theta =0 \label{angle}
\end{eqnarray}
where $\triangle=r^2-2Mr+a^2$, $K=(r^2+a^2)\omega -am$ and
$\bar{\lambda} =E+a^2\omega^2-2am\omega $. The parameter $s$, the
spin-weight of the perturbation fields could be $s=0, \pm\frac12,
\pm1, \pm2$, and corresponds to the scalar, neutrino,
electromagnetic or gravitational perturbations respectively.
Eq.(\ref{angle}) is the spin-weighted spheroidal wave equation. This
is kind of the Sturm-liuvelle problem, and the boundary conditions
requires $\Theta$ is finite at $\theta=0,\ \pi$. Though it has been
widely used in many fields,  it has not been well studied due to the
mathematical difficulty[1-5]. Until recent years, with the
introduction of super-symmetric quantum mechanics to study the
spheroidal equations (that is, s=0 case), it becomes possible to
solve it[6-12]. This article would apply this method to study it in
the case of $s=1/2$. When $s=0$, the problem is more complex than
previous work. So more calculation is need in determining the
eigenvalue.
\section{Calculation of the first several terms of the super-potential}
  In order to employ the super-symmetry quantum mechanics (SUSYQM) to Eq.(\ref{angle}),
   we first make it into the
Schr\"{o}dinger form[6-12]
\begin{eqnarray}
\frac{d^2\Psi}{d\theta^2}&+&\bigg[\frac{1}{4}+s+\beta^{2}\cos ^2
\theta -2s\beta \cos\theta\nonumber\\
&-&\frac{(m+s\cos\theta)^2-\frac{1}{4}}{\sin ^2\theta}+E\bigg]\Psi=0
\label{new eq}
\end{eqnarray}
by the transformation
\begin{eqnarray}
\Theta(\theta)=\frac{\Psi(\theta)}{\sqrt{sin\theta}}
\end{eqnarray}
 with $\beta=a\omega$.  The corresponding  boundary conditions now
 become
$\Psi|_{\theta=0}=\Psi|_{\theta=\pi}=0$. For Eq.(\ref{new eq}), we
make a few instructions: Eq.(\ref{new eq}) just has the
Schr\"{o}dinger form, it does not represent a particular quantum
system. For convenience, we apply the terminology of quantum
mechanics, such as potential energy, ground state, excited state,
etc in the following. The potential in Eq.(\ref{new eq}) is
\begin{eqnarray}
V(\theta,\beta,s)&=&-\bigg[\frac{1}{4}+s+\beta^{2}\cos ^2 \theta
-2s\beta \cos\theta\nonumber\\
&&+\frac{(m+s\cos\theta)^2-\frac{1}{4}}{\sin
^2\theta}\bigg]\label{potential},
\end{eqnarray}
Super potential $W$ is a core concept in SUSYQM,  and it is
connected with the potential by[13]
\begin{equation}
W^2-W'=V(\theta,\beta,s)-E_0.\label{potential and super relation}
\end{equation}
According to the theory of the SUSYQM, the form of ground
eigenfunction  $\Psi_0$ is completely known through the
super-potential $W$ by the formula [13]
\begin{eqnarray}
\Psi_0&=&N\exp\bigg[-\int Wd\theta\bigg]\label{psi}.
\end{eqnarray}
Hence, the key work in SUSYQM is solve Eq.(\ref{potential and super
relation}). Practically, Eq.(\ref{potential and super relation}) is
the same difficult to deal with as Eq.(\ref{new eq}). Therefore, we
rely on the perturbation methods to treat it. That is, we study it
by  expanding the super-potential $W$ and the ground eigenvalue
$E_0$ into series form of  the parameter $\beta$ [6-12]:
\begin{equation}
W=\sum_{n=0}^{\infty}\beta^nW_{n}\label{super-potential expansion}
\end{equation}
\begin{eqnarray}
E_0=\sum_{n=0}^{\infty}E_{0,n;m}\beta^n\label{energy expansion},
\end{eqnarray}
where the lower suffix $m$ in $E_{0,n;m}$ is referred to the
parameter $m$ in Eq.(\ref{new eq}) and the index $0$ means belonging
to the ground state energy and $n$ refers to the nth-order item of
the series expansion.

Substituting Eqs.(\ref{super-potential expansion}), Eq.(\ref{energy
expansion}) into Eq.(\ref{potential and super relation}) gives the
following equations[11]
\begin{eqnarray}
W'_0-W_0^2&=&E_{0,0;m}+s+\frac14\nonumber\\
&&-\frac{(m+s\cos\theta)^2-\frac{1}{4}}{\sin ^2\theta}\equiv f_0(\theta)\label{w0equation}\\
W'_1-2W_0W_1&=&-2s\cos\theta +E_{0,1;m}\nonumber\\ & \equiv &f_1(\theta)\\
W'_2-2W_0W_2&=&\cos^2\theta+W_1^2+E_{0,2;m}\nonumber\\ & \equiv & f_2(\theta)\\
W'_{n}-2W_{0}W_{n}&=&\sum_{k=1}^{n-1}W_{k}W_{n-k}+E_{0,n;m}\nonumber\\
& \equiv & f_n(\theta),\  (n\geq3)\label{fn}
\end{eqnarray}
by equating  the coefficients of the same power of $\beta$ in its
two sides. Our main work is to find the solutions of the above
equations in the case of $s=\frac12$.

 The solution of Eq.(\ref{w0equation}) is easy to find [11]
\begin{eqnarray}
E_{0,0;m}&=&m^2+m-3/4\nonumber\\
W_0&=&-\frac{1/2+(m+\frac12)\cos\theta}{sin\theta}\label{W_0}
\end{eqnarray}
With $W_0$ known, it is easy to give $W_n$ on according to the
knowledge of differential equations,
\begin{eqnarray}
W_{n}(\theta)&=&e^{2\int W_{0}d\theta}A_{n}(\theta)\nonumber\\
&=&\bigg[\tan\frac{\theta}2 \sin^{2m+1}\theta\bigg]^{-1}
A_{n}(\theta) \label{A-n,W-n relation}
\end{eqnarray}
where,
\begin{eqnarray}
A_{n}(\theta)&=& \int f_n(\theta)e^{-2\int W_{0}d\theta}
d\theta\nonumber\\ &=&\int f_n(\theta)\tan\frac{\theta}2
\sin^{2m+1}\theta d\theta \label{A-n eqn}.
\end{eqnarray}

In order to calculate Eqs.(\ref{A-n,W-n relation})-(\ref{A-n eqn}),
we needs the following integral formulae [14]:
\begin{eqnarray}
&&P(2m,\theta)\nonumber\\
&=&\int{sin^{2m}\theta}d\theta\nonumber\\
&=&-\frac{\cot\theta}{2m+1}\bigg[\sum_{k=0}^{m-1}\bar{I}(2m,k)sin^{2m-2k}\theta\bigg]+\frac{(2m-1)!!}{(2m)!!}\theta\label{defination
P}\label{p2m}
\end{eqnarray}
where
\begin{eqnarray}
\bar{I}(2m,k)&=&\frac{(2m+1)(2m-1)\cdot\cdot\cdot(2m-2k+1)}{2m(2m-2)\cdot\cdot\cdot(2m-2k)},\nonumber\\
&& \ \ \ \ \ \ \ \ \ \ \ \ \ \ \ \ \ \ \ \ \   (k\geq0)
\end{eqnarray}
After a simple derivation, we get
\begin{eqnarray}
&&P(2m+2,\theta)\nonumber\\
&=&\frac{2m+1}{2m+2}P(2m,\theta)-\frac{\cos\theta\sin^{2m+1}\theta}{2m+2},
\end{eqnarray}
\begin{eqnarray}
&&P(2m+4,\theta)\nonumber\\
&=&\frac{(2m+1)(2m+3)}{(2m+2)(2m+4)}P(2m,\theta)\nonumber\\
&&-\frac{\cos\theta\sin^{2m+3}}{2m+4}-\frac{(2m+3)\cos\theta\sin^{2m+1}\theta}{(2m+2)(2m+4)}\\
&&P(2m+6,\theta)\nonumber\\
&=&\frac{(2m+1)(2m+3)(2m+5)}{(2m+2)(2m+4)(2m+6)}P(2m,\theta)\nonumber\\
&&-\frac{\cos\theta\sin^{2m+5}}{2m+6}
-\frac{(2m+5)\cos\theta\sin^{2m+3}\theta}{(2m+4)(2m+6)}\nonumber\\
&&-\frac{(2m+3)(2m+5)\cos\theta\sin^{2m+1}\theta}{(2m+2)(2m+4)(2m+6)}
\end{eqnarray}
By the above equation,we can summarize the general formula
\begin{eqnarray}
&&P(2m+2n)\nonumber\\
&=&\frac{2m+1}{2m+2n+1} \bar{I}(2m+2n,n-1)P(2m,\theta)\nonumber\\
&&-\cos\theta\sum_{l=1}^{n}\frac{\bar{I}(2m+2n,n-l)}{2m+2n+1}\sin^{2m+2l-1}\theta\label{Integral
formula}
\end{eqnarray}
This formula can be proved by mathematical induction. By the help of
Eqs.(\ref{W_0}), (\ref{Integral formula}), $A_{1}(\theta)$ is now
simplified as
\begin{eqnarray}
A_{1}(\theta)&=&\int (E_{0,1;m}-cos\theta)(1-cos\theta)sin^{2m}\theta d\theta\nonumber\\
     &=&(E_{0,1;m}+1)P(2m,\theta)-P(2m+2,\theta)\nonumber\\ &&-\frac{E_{0,1;m}+1}{2m+1}\sin^{2m+1}\theta\nonumber\\
     &=&[E_{0,1;m}+1-\frac{2m+1}{2m+2}]P(2m,\theta)\nonumber\\ &&+\bigg[\frac{\cos\theta}{2m+2}-\frac{E_{0,1;m}+1}{2m+1}\bigg]\sin^{2m+1}\theta
\end{eqnarray}
Now we discuss the term $P(2m,\theta)$. According to Eq.(\ref{psi})
and Eq.(\ref{p2m}), we can see that $\Psi_0(\theta)$ is $\infty$ at
the boundaries $\theta=0,\ \pi$. This result does not meet the
boundary conditions that $\Psi(\theta)$ should finite at $\theta=0,\
\pi$. So that the coefficient of the term $P(2m,\theta)$ must be
zero. Thus,
\begin{equation}
E_{0,1}=-\frac{1}{2m+2}\label{E-1}
\end{equation}
Now we can simplify $A_1$
\begin{eqnarray}
A_{1}(\theta)&=&-\frac{sin^{2m+1}\theta}{2m+2}+\frac{cos\theta
sin^{2m+1}\theta}{2m+2}
\end{eqnarray}
With the help of Eq.(\ref{A-n,W-n relation}) and Eq.(\ref{W_0}), it
is easy to obtain the first order $W_1(\theta)$
\begin{eqnarray}
W_{1}(\theta)&=&-\frac{1}{2m+2}\sin\theta
\end{eqnarray}
By the same tedious calculation as that of $W_1(\theta)$ ,
$E_{0,2;m}-E_{0,4;m}$ and $W_2(\theta)-W_4(\theta)$can also be
obtained. The results are
\begin{eqnarray}
E_{0,2;m}&=&-\frac{4m^2+10m-5}{(2m+2)^3}\\
E_{0,3;m}&=&-\frac{4(2m+1)^{2}(2m+3)}{(2m+2)^{5}(2m+4)}\\
E_{0,4;m}&=&-\frac{2(2m+1)^{2}(2m+3)(2m^2+9m+2)}{(2m+2)^{7}(2m+4)}\nonumber\\
W_2(\theta)&=&b_{2,1}sin\theta+a_{2,1}sin\theta\cos\theta\\
W_{3}(\theta)&=&b_{3,1}\sin\theta+b_{3,2}\sin^{3}\theta+a_{3,1}\sin\theta\cos\theta\\
W_{4}(\theta)&=&b_{4,1}\sin\theta+b_{4,1}\sin^{3}\theta\nonumber\\
&&+a_{4,1}\sin\theta\cos\theta+a_{4,2}\sin^{3}\theta\cos\theta
\end{eqnarray}
Where,
\begin{eqnarray}
b_{2,1}&=&-\frac{2m+1}{(2m+2)^3}\ \ a_{2,1}=\frac{2m+1}{(2m+2)^2}\\
b_{3,1}&=&\frac{4(2m+1)}{(2m+2)^{5}(2m+4)}\\
b_{3,2}&=&-\frac{2(2m+1)}{(2m+2)^{3}(2m+4)}\\
a_{3,1}&=&-\frac{4(2m+1)}{(2m+2)^{4}(2m+4)}\\
b_{4,1}&=&\frac{2(2m+1)(2m^2+9m+2)}{(2m+2)^{7}(2m+4)}\\
a_{4,1}&=&-\frac{2(2m+1)(2m^2+9m+2)}{(2m+2)^{6}(2m+4)}\\
b_{4,2}&=&-\frac{6m(2m+1)}{(2m+2)^{5}(2m+4)}\\
a_{4,2}&=&\frac{2m(2m+1)}{(2m+2)^{4}(2m+4)}
\end{eqnarray}

\section{Calculation of the general n-th terms of the super-potential}
From the four terms of $W_1-W_4$, we hypothetically summarize a
general formula for $W_n$ as
\begin{equation}
W_{n}(\theta)=\sum_{k=1}^{[\frac{n}{2}]}a_{n,k}\sin^{2k-1}\theta\cos\theta+\sum_{k=1}^{[\frac{n+1}{2}]}b_{n,k}\sin^{2k-1}\theta\label{w_n}
\end{equation}
Here we use mathematical induction to prove that the guess is true.

First it is easy to see the assumption (\ref{w_n}) is the same as
that of $W_1$ when $N=1$. Under the condition that all $W_N$ meet
the requirement of (\ref{w_n}) whenever $N\le n-1$, we will try to
solve the differential equation for $W_n$ to verify that it also can
be written as that of (\ref{w_n}). Back to
Eqs.(\ref{fn}),(\ref{A-n,W-n relation}),(\ref{A-n eqn}), one needs
to simplify the term $\sum_{k=1}^{n-1}W_{k}W_{n-k}$ in order to
calculate $W_n$. Whenever $1\le k \le n-1$, one has $1\le n-k\le
n-1$ and $W_k(\theta)$, $W_{n-k}(\theta)$ could be written in the
form of (\ref{w_n}). That is,
\begin{eqnarray}
W_{k}(\theta)&=&\sum_{i=1}^{[\frac{k}{2}]}a_{k,i}\sin^{2i-1}\theta\cos\theta+\sum_{i=1}^{[\frac{k+1}{2}]}b_{k,i}\sin^{2i-1}\theta\nonumber\\
W_{n-k}(\theta)&=&\sum_{j=1}^{[\frac{n-k}{2}]}a_{n-k,j}\sin^{2j-1}\theta\cos\theta\nonumber\\
&&+\sum_{j=1}^{[\frac{n-k+1}{2}]}b_{n-k,j}\sin^{2j-1}\theta
\end{eqnarray}
For the sake of later use, the facts are true
\begin{eqnarray}
a_{i,j}=0, j<1 \ or\  j>[\frac{i}2];\nonumber\\ b_{i,j}=0, \ j<1 \
or\ j>[\frac{i+1}2]\label{a-[i,j]=0 condition}
\end{eqnarray}
for $i<n$.  Thus
\begin{eqnarray}
\sum_{k=1}^{n-1}W_{k}W_{n-k}=\bar{A}_1+\bar{A}_2+\bar{A}_3+\bar{A}_4
\end{eqnarray}
where
\begin{eqnarray}
\bar{A}_1&=&\sum_{k=1}^{n-1}\sum_{i=1}^{[\frac{k+1}{2}]}\sum_{j=1}^{[\frac{n-k+1}{2}]}b_{k,i}b_{n-k,j}\sin^{2i+2j-2}\theta\nonumber\\
\bar{A}_2&=&\sum_{k=1}^{n-1}\sum_{i=1}^{[\frac{k}{2}]}\sum_{j=1}^{[\frac{n-k}{2}]}a_{k,i}b_{n-k,j}\sin^{2i+2j-2}\theta\cos^{2}\theta\nonumber\\
\bar{A}_3&=&\cos\theta\sum_{k=1}^{n-1}\bigg[\sum_{i=1}^{[\frac{k}{2}]}\sum_{j=1}^{[\frac{n-k+1}{2}]}a_{k,i}b_{n-k,j}\sin^{2i+2j-2}\theta\bigg]\nonumber\\
\bar{A}_4&=&\cos\theta\sum_{k=1}^{n-1}\bigg[\sum_{i=1}^{[\frac{k+1}{2}]}\sum_{j=1}^{[\frac{n-k}{2}]}b_{k,i}a_{n-k,j}\sin^{2i+2j-2}\theta\bigg]
\end{eqnarray}
exchanging the order of the sums in all the above equations, they
are simplified as
\begin{eqnarray}
\bar{A}_1&=&\sum_{p=2}^{\bar{c}_1}\sum_{k=1}^{n-1}\sum_{j=1}^{p-1}b_{k,p-j}b_{n-k,j}sin^{2p-2}\theta\\
\bar{A}_2&=&\sum_{p=2}^{\bar{c}_2-1}\sum_{k=1}^{n-1}\sum_{j=1}^{p-1}a_{k,p-j}a_{n-k,j}sin^{2p-2}\theta\nonumber\\
&& -\sum_{p=3}^{\bar{c}_2}\sum_{k=1}^{n-1}\sum_{j=1}^{p-1}a_{k,p-1-j}a_{n-k,j}sin^{2p-2}\theta\nonumber\\
\bar{A}_3&=&+\cos\theta\sum_{p=2}^{\bar{c}_3}\sum_{k=1}^{n-1}\sum_{j=1}^{p-1}b_{k,p-j}a_{n-k,j}sin^{2p+1}\theta\nonumber\\
\bar{A}_4&=&\cos\theta\sum_{p=2}^{\bar{c}_4}\sum_{k=1}^{n-1}\sum_{j=1}^{p-1}a_{k,p-j}b_{n-k,j}sin^{2p+1}\theta
\end{eqnarray}
where\begin{eqnarray}
 \bar{c}_1&=& \bigg[\frac{k+1}{2}\bigg]+\bigg[\frac{n-k+1}{2}\bigg],\ \bar{c}_2=\bigg[\frac{k}{2}\bigg]+\bigg[\frac{n-k}{2}\bigg]+1\nonumber\\
 \bar{c}_3&=& \bigg[\frac{k+1}{2}\bigg]+\bigg[\frac{n-k}{2}\bigg],\ \bar{c}_4=\bigg[\frac{k}{2}\bigg]+\bigg[\frac{n-k+1}{2}\bigg]
\end{eqnarray}
It is easy to see \begin{eqnarray}
 \bar{c}_1=\bar{c}_2=\frac n2+1,\  \ \bar{c}_3=\bar{c}_4=\frac n2, \ when\  n\  is\  even \\
 \ \bar{c}_1=  \bar{c}_2=\bar{c}_3=
 \bar{c}_4=\frac{n+1}2,\ \ when\  n\  is\ odd
\end{eqnarray}

Hence
\begin{eqnarray}\sum_{k=1}^{n-1}W_{k}W_{n-k}=\sum_{p=2}^{[\frac{n}{2}]+1}\bigg[h_{n,p}+g_{n,p}\cos\theta\bigg]sin^{2p-2}\theta\label{wk-wn-k}
\end{eqnarray}
where  $h_{n,p}$ and $h_{n,p}$ are constant coefficients:
\begin{eqnarray}
h_{n,p}=\sum_{k=1}^{n-1}\sum_{j=1}^{p-1}\bigg[b_{k,p-j}b_{n-k,j}+a_{k,p-j}a_{n-k,j}
-a_{k,p-1-j}a_{n-k,j}\bigg]\\
 g_{n,p}=\sum_{k=1}^{n-1}\sum_{j=1}^{p-1}\bigg[b_{k,p-j}a_{n-k,j}+a_{k,p-j}b_{n-k,j}\bigg]\end{eqnarray}

Hence, one has
\begin{eqnarray}
\ \ when\  n\  is\  even \nonumber\\
 g_{n,p}&=&0,\  \ p<2\  or\  p> \frac n2; \nonumber\\
  h_{n,p}&=&0,\  \ p<2\ or\  p> \frac n2+1,\nonumber\\
\ \ when\  n\  is\ odd \nonumber\\
 g_{n,p}&=&h_{n,p}=0,\  \ p<2\ or\ p> \frac{n+1}2, \label{g-np,h-np=0 condition}.
\end{eqnarray}
 We have  used  the fact (\ref{a-[i,j]=0
condition}) and substituted the quantities $\bar{c}_1,\ \bar{c}_2,\
\bar{c}_3,\ \bar{c}_4$ by the maximum $[\frac{n}{2}]+1$ of them in
last line in the above equation for convenience in the following
calculation process. On the final conclusion, we will take
Eq.(\ref{g-np,h-np=0 condition}) into consideration. Substituting
$f_n(z)=\sum_{k=1}^{n-1}W_{k}W_{n-k}+E_{0,n;m}$ into Eqs.(\ref{A-n
eqn}) and by the use of Eq.(\ref{p2m}), we can have
\begin{eqnarray}
&&A_{n}(\theta)\nonumber\\
&=&\int\sum_{p=2}^{[\frac{n}{2}]+1}\bigg[h_{n,p}+g_{n,p}\cos\theta\bigg]sin^{2p-2}\theta(1-cos\theta)sin^{2m}\theta d\theta\nonumber\\
&&+\int E_{0,n;m}*(1-cos\theta)sin^{2m}\theta d\theta\nonumber\\
&=&\sum_{p=2}^{[\frac{n}{2}]+1}\frac{g_{n,p}-h_{n,p}}{2m+2p-1}sin^{2m+2p-1}\theta-\frac{E_{0,n;m}}{2m+1}\sin^{2m+1}\theta\nonumber\\
&&+\sum_{p=2}^{[\frac{n}{2}]+1}\bigg(h_{n,p}-g_{n,p}\bigg)P(2m+2p-2,\theta)\nonumber\\
&&+\sum_{p=2}^{[\frac{n}{2}]+1}g_{n,p}P(2m+2p,\theta)+E_{0,n;m}P(2m,\theta)
\end{eqnarray}
According to Eq.(\ref{Integral formula}), one has
\begin{eqnarray}
P(2m+2p-2)&=&\frac{2m+2p}{2m+2p-1}P(2m+2p,\theta)\nonumber\\
&&-\cos\theta\frac{\sin^{2m+2p-1}\theta}{2m+2p-1}\end{eqnarray} and
\begin{eqnarray}
&&P(2m+2p)\nonumber\\
&=&\frac{(2m+1)\bar{I}(2m+2p,p-1)}{2m+2p+1}P(2m,\theta)\nonumber\\
&&
-\cos\theta\sum_{l=1}^{p}\frac{\bar{I}(2m+2p,p-l)}{2m+2p+1}\sin^{2m+2l-1}\theta
\label{Integral formula n=p-1}
\end{eqnarray}
Hence,
\begin{eqnarray}
&&A_{n}(\theta)\nonumber\\
&=&\sum_{p=2}^{[\frac{n}{2}]+1}\frac{g_{n,p}-h_{n,p}}{2m+2p-1}sin^{2m+2p-1}\theta-\frac{E_{0,n;m}}{2m+1}\sin^{2m+1}\theta\nonumber\\
&&-\cos\theta\sum_{p=2}^{[\frac{n}{2}]+1}\frac{g_{n,p}-h_{n,p}}{2m+2p-1}sin^{2m+2p-1}\theta\nonumber\\
&&-\cos\theta \sum_{p=2}^{[\frac{n}{2}]+1}e_{n,p}\ast
\sum_{l=1}^{p}\frac{\bar{I}(2m+2p,p-l)}{2m+2p+1}\sin^{2m+2l-1}\theta\nonumber\\
&&+b *P(2m,\theta)\label{A-n-1}
\end{eqnarray}
where the quantities $c_{n,p}$, and $ b$ as the coefficient of
$P(2m,\theta )$ are
\begin{eqnarray}
&&e_{n,p}=\frac{2m+2p}{2m+2p-1}(h_{n,p}-\frac{g_{n,p}}{2m+2p}),\\
&&b=E_{0,n;m}+\sum_{p=2}^{[\frac{n}{2}]+1}e_{n,p}\ast\frac{(2m+1)\bar{I}(2m+2p,p-1)}{2m+2p+1}
\end{eqnarray}
respectively. As stated before, the coefficient $b$ must be zero to
make the eigenfunction finite at the boundary, then
\begin{eqnarray}
E_{0,n;m}&=&-\sum_{p=2}^{[\frac{n}{2}]+1}e_{n,p}\ast\frac{(2m+1)\bar{I}(2m+2p,p-1)}{2m+2p+1}\label{e0nm}
\end{eqnarray}
Now, we first simplify some terms in Eq.(\ref{A-n-1})
\begin{eqnarray}
&& \cos\theta \sum_{p=2}^{[\frac{n}{2}]+1}e_{n,p}\ast
\sum_{l=1}^{p}\frac{\bar{I}(2m+2p,p-l)}{2m+2p+1}\sin^{2m+2l-1}\theta\nonumber\\
&=&\cos\theta \sum_{p=2}^{[\frac{n}{2}]+1}e_{n,p}\ast
\sum_{l=2}^{p}\frac{\bar{I}(2m+2p,p-l)}{2m+2p+1}\sin^{2m+2l-1}\theta\nonumber\\
&&+\cos\theta\sum_{p=2}^{[\frac{n}{2}]+1}e_{n,p}\frac{\bar{I}(2m+2p,p-1)}{2m+2p+1}\sin^{2m+1}\theta\nonumber\\
&=&\cos\theta\sum_{l=2}^{[\frac{n}{2}]+1}\sum_{p=l}^{[\frac{n}{2}]+1}e_{n,p}\frac{\bar{I}(2m+2p,p-l)}{2m+2p+1}\sin^{2m+2l-1}\theta\nonumber\\
&&+\cos\theta\sum_{p=2}^{[\frac{n}{2}]+1}e_{n,p}\frac{\bar{I}(2m+2p,p-1)}{2m+2p+1}\sin^{2m+1}\theta
\label{A-n simplifying}
\end{eqnarray}
Taking Eqs.(\ref{e0nm}), (\ref{A-n simplifying}) into
Eq.(\ref{A-n-1}),we can obtain
\begin{eqnarray}
&&A_{n}(\theta)\nonumber\\
&=&\sum_{p=2}^{[\frac{n}{2}]+1}\frac{g_{n,p}-h_{n,p}}{2m+2p-1}sin^{2m+2p-1}\theta-\frac{E_{0,n;m}}{2m+1}\sin^{2m+1}\theta\nonumber\\
&&-\cos\theta\sum_{p=2}^{[\frac{n}{2}]+1}\frac{g_{n,p}-h_{n,p}}{2m+2p-1}sin^{2m+2p-1}\theta\nonumber\\
&&-\cos\theta\sum_{l=2}^{[\frac{n}{2}]+1}\sum_{p=l}^{[\frac{n}{2}]+1}e_{n,p}\frac{\bar{I}(2m+2p,p-l)}{2m+2p+1}\sin^{2m+2l-1}\theta\nonumber\\
&&-\cos\theta\sum_{p=2}^{[\frac{n}{2}]+1}e_{n,p}\frac{\bar{I}(2m+2p,p-1)}{2m+2p+1}\sin^{2m+1}\theta\nonumber\\
&=&F_1+\cos\theta F_2\label{A-n-2}
\end{eqnarray}
 where $F_1,\ F_2$ are
\begin{eqnarray}
F_1=\sum_{p=2}^{[\frac{n}{2}]+1}\frac{g_{n,p}-h_{n,p}}{2m+2p-1}sin^{2m+2p-1}\theta-\frac{E_{0,n;m}}{2m+1}\sin^{2m+1}\theta\end{eqnarray}
\begin{eqnarray}
F_2&=&-\sum_{p=2}^{[\frac{n}{2}]+1}\frac{g_{n,p}-h_{n,p}}{2m+2p-1}sin^{2m+2p-1}\theta\nonumber\\
&&-\sum_{l=1}^{[\frac{n}{2}]+1}\sum_{p=l}^{[\frac{n}{2}]+1}e_{n,p}\frac{\bar{I}(2m+2p,p-l)}{2m+2p+1}\sin^{2m+2l-1}\theta\nonumber\\
&=&-\sum_{p=2}^{[\frac{n}{2}]+1}\frac{g_{n,p}-h_{n,p}}{2m+2p-1}sin^{2m+2p-1}\theta\nonumber\\
&&-\sum_{p=2}^{[\frac{n}{2}]+1}e_{n,p}\frac{\bar{I}(2m+2p,p-1)}{2m+2p+1}\sin^{2m+1}\theta\nonumber\\
&&-\sum_{l=2}^{[\frac{n}{2}]+1}\sum_{p=l}^{[\frac{n}{2}]+1}e_{n,p}\frac{\bar{I}(2m+2p,p-l)}{2m+2p+1}\sin^{2m+2l-1}\theta\nonumber
\end{eqnarray}
With the help of Eq.(\ref{A-n,W-n relation}),  it is easy to obtain
\begin{eqnarray}
&&W_n\nonumber\\
&=&\frac{\cos\theta\bigg[F_1+F_2\bigg]+\bigg[F_1+F_2-\sin^2\theta
F_2\bigg]}{\sin^{2m+2}\theta}\nonumber\\
&=&\frac{\cos\theta\bigg[F_1+F_2\bigg]}{\sin^{2m+2}\theta}\nonumber\\
&&+\frac{F_1+F_2-\sin^2\theta(F_1+F_2)+F_1\sin^2\theta}{\sin^{2m+2}\theta}.
\end{eqnarray}
With the help of Eq.(\ref{e0nm}), it is easy to have
\begin{eqnarray}
F_1+F_2=-\sum_{l=2}^{[\frac{n}{2}]+1}\sum_{p=l}^{[\frac{n}{2}]+1}e_{n,p}\frac{\bar{I}(2m+2p,p-l)}{2m+2p+1}\sin^{2m+2l-1}\theta.
\end{eqnarray}
  one achieves
\begin{eqnarray}
W_n&=&\cos\theta
\sum_{l=2}^{[\frac{n}{2}]+1}a_{n,l-1}\sin^{2l-3}\theta+\sum_{l=2}^{[\frac{n}{2}]+1}b_{n,l-1}\sin^{2l-3}\theta.
\label{wn to be proved}
\end{eqnarray}
Where,
\begin{eqnarray}
a_{n,l-1}&=&-\sum_{p=l}^{[\frac{n}{2}]+1}e_{n,p}\frac{\bar{I}(2m+2p,p-l)}{2m+2p+1},
\ l\ge 2\nonumber\\
&=&-\sum_{p=l}^{[\frac{n}{2}]+1}e_{n,p}\frac{(2m+2l-2)\bar{I}(2m+2p,p-l+1)}{(2m+2l-1)(2m+2p+1)}
\\b_{n,1}&=&a_{n,1}-\frac{E_{0,n;m}}{2m+1}\\
b_{n,l-1}
&=&a_{n,l-1}-a_{n,l-2}+\frac{g_{n,l-1}-h_{n,l-1}}{2m+2l-3},\ l\ge
3,\label{b-n,l--a-n,l+1}
\end{eqnarray}
Eq.(\ref{wn to be proved}) could be written as
\begin{eqnarray}
W_n=\cos\theta
\sum_{l=1}^{[\frac{n}{2}]}a_{n,l}\sin^{2l-1}\theta+\sum_{l=1}^{[\frac{n}{2}]+1}b_{n,l}\sin^{2l-1}\theta
\label{wn to be proved 0}
\end{eqnarray}
where for $l\ge 1$
\begin{eqnarray}
a_{n,l}&=&-\sum_{p=l+1}^{[\frac{n}{2}]+1}e_{n,p}\frac{(2m+2l)\bar{I}(2m+2p,p-l)}{(2m+2l+1)(2m+2p+1)}\label{a-n,l--a-n,l},
\\ b_{n,1}&=&a_{n,1}-\frac{E_{0,n;m}}{2m+1}\\
b_{n,l}&=&a_{n,l}-a_{n,l-1}+\frac{g_{n,l}-h_{n,l}}{2m+2l-1},\ l\ge
2.\label{b-n,l--a-n,l}
\end{eqnarray}
Compare Eqs.(\ref{wn to be proved 0}), (\ref{w_n}), their difference
lies in the upper limit of the second sum in $W_n$
 Whenever $n$ is odd,  $[\frac{n+1}2]=[\frac n2]+1$ means Eq. (\ref{w_n}) the same as Eq.(\ref{wn to be proved 0}). While $n$ is
 even, we could use Eq. (\ref{a-[i,j]=0 condition}) and Eq. (\ref{g-np,h-np=0 condition}) to obtain
\begin{eqnarray}
a_{n,\frac n2+1}&=&0\\ b_{n,[\frac
n2]+1}&=&b_{n,\frac n2+1}\nonumber\\
&=&-a_{n,\frac n2}-\frac{h_{n,\frac n2+1}}{2m+n+1}=0
\end{eqnarray}
Hence the upper limit for $b_{n,p}$ becomes $\frac n2=[\frac{n+1}2]$
when $n$ is even. The correctness of our induction about the general
formula with $W_n$ is completed. According to
Eq.(\ref{a-n,l--a-n,l}),Eq.(\ref{b-n,l--a-n,l}) and
Eq.(\ref{e0nm}),we can get some interesting results:
\begin{eqnarray}
b_{n,l}&=&\frac{1}{2m+2l}(g_{n,l}-a_{n,l}),\ l\ge 2 \label{b_nl}
\end{eqnarray}
\begin{eqnarray}
a_{n,1}&=&\frac{2m+2}{(2m+1)(2m+3)}E_{0,n;m}
\end{eqnarray}
We can compare these interesting results with $W_3(\theta)$ and
$W_4(\theta)$ to verify the correctness of the general formula of
$W_{n}(\theta)$.The validation results are satisfactory. So that,we
can say that the general formula of $W_{n}(\theta)$ is accurate.

\section{The ground  state eigenfunctions}
Based on the above conclusions, the super-potential $W$ could be
written as
\begin{eqnarray}
&&W=W_0+\sum_{n=1}^{\infty}\beta^nW_n
\end{eqnarray}
The ground eigenfunction becomes
\begin{eqnarray}
\Psi_0&=&N(1-\cos\theta)^{\frac{1}{2}}\sin^{m}\theta\nonumber\\
&&\ast\exp\bigg[-\sum_{n=1}^{\infty}\beta^n
        \bigg(\sum_{k=1}^{[\frac{n}{2}]}a_{n,k}\frac{\sin^{2k}\theta}{2k}\nonumber\\
        &&+\sum_{k=1}^{[\frac{n+1}{2}]}b_{n,k}P(2k-1,\theta)\bigg)\bigg]
\end{eqnarray}
Back to the $\Theta$, the above ground eigenfunction becomes
\begin{eqnarray}
\Theta_0&=&N(1-\cos\theta)^{\frac{1}{2}}\sin^{m-\frac{1}{2}}\theta\nonumber\\
&&\ast\exp\bigg[-\sum_{n=1}^{\infty}\beta^n
        \bigg(\sum_{k=1}^{[\frac{n}{2}]}a_{n,k}\frac{\sin^{2k}\theta}{2k}\nonumber\\
        &&+\sum_{k=1}^{[\frac{n+1}{2}]}b_{n,k}P(2k-1,\theta)bigg)\bigg]
\end{eqnarray}
and the ground energy is
\begin{eqnarray}
E_{0;m}=m(m+1)-\frac{3}{4}+\sum_{n=1}^{\infty}E_{0,n;m}\beta^n
\end{eqnarray}
with $E_{0,n;m}$ determined by Eq.(\ref{e0nm}).

\section{The excited eigenfunctions  }
In the following, we will compute the excited eigenfunctions. As
done in Ref.\cite{Li5},  we hope to extend the study of the
recurrence relations by the means of super-symmetric quantum
mechanics to Eq. (\ref{new eq}).

The super-potential $W$  connects the two partner potential
$V_{\mp}$ by
\begin{equation} \label{vpm} V^{\mp}(\theta)=W^2(\theta) \mp W'(\theta).
\end{equation}
The shape-invariance properties mean that the pair of partner
potentials $V^{\pm}(x)$  are similar in shape and differ only in the
parameters, that is
\begin{equation} \label{shape invariant} V^+(\theta;a_1) = V^-(\theta;a_2) +
R(a_1),
\end{equation}
where $a_1$ is a set of parameters, $a_2$ is a function of $a_1$
(say $a_2=f(a_1)$) and the remainder $R(a_1)$ is independent of
$\theta$.

We must introduce the parameters $A_{i,j},\ B_{i,j}$ into the
super-potential $W$ in order to study the shape-invariance
properties of the spin-weighted spheroidal equations as:
\begin{eqnarray}
W(A_{n,j},B_{n,j},\theta)&=&-A_{0,0}(m+\frac12)\cot
\theta-\frac{1}{2}B_{0,0}\csc\theta \nonumber\\ &&
+\sum_{n=1}^{\infty}\beta^nW_n(A_{n,j},B_{n,j},\theta),
\end{eqnarray}
where
\begin{eqnarray}
&&W_n(A_{n,j},B_{n,j},\theta)\nonumber\\
&& \ \ \ \
=\sum_{j=1}^{[\frac{n+1}{2}]}\bar{b}_{n,j}\sin^{2j-1}\theta +\cos
\theta\sum_{j=1}^{[\frac{n}{2}]}\bar{a}_{n,j}\sin^{2j-1}\theta
\end{eqnarray}
with
\begin{eqnarray}
\bar{a}_{n,j} =A_{n,j}a_{n,j},\  \bar{b}_{n,j} =B_{n,j}b_{n,j}
\end{eqnarray}
Then, $V^{\pm}(A_{n,j},B_{n,j},\theta)$ are
$V^{\pm}(A_{n,j},B_{n,j},\theta)$ are defined as
\begin{eqnarray}
V^{\pm}(A_{n,j},B_{n,j},\theta)=W^2(A_{n,j},B_{n,j},\theta)\pm W'\nonumber\\
=\sum_{n=0}^{\infty}\beta^nV^{\pm}_n(A_{i,j},B_{n,j},\theta).
\end{eqnarray}
The key point is to try to find some quantities $C_{i,j},D_{i,j}$ to
make the relations
\begin{eqnarray}
V^{+}_n(A_{i,j},B_{n,j},\theta)=V^{-}_n(C_{i,j},D_{n,j},\theta)+R_{n;m}(A_{i,j},B_{n,j})\label{shape-invariance
v-pm }
\end{eqnarray} retain with  $R_{n;m}(A_{i,j},B_{n,j})=R_{n;m}$ pure quantities.
Now, we will prove that it is true for the special cases $n=0,\
1,2$. It is easy to obtain
\begin{eqnarray}
V^+_0(A_{0,0},B_{0,0},\theta) &=& V^-_0(C_{0,0},D_{0,0},\theta)+R_{0;m}\\
V^+_1( A_{1,1},B_{1,1}, \theta)&=& V^-_1( C_{1,1},D_{1,1},
\theta)+R_{1;m}\\
V^+_2( B_{2,1},A_{2,1}, \theta)&=& V^-_2( C_{2,1},D_{2,1},
\theta)+R_{2;m}.
\end{eqnarray}
In order to retain the above equations, one must have
\begin{eqnarray}
C_{0,0}&=&A_{0,0}+\frac2{2m+1},\\ D_{0,0}&=&B_{0,0}\\
D_{1,1}&=&\frac{(2m+1)A_{0,0}-1}{(2m+1)A_{0,0}+3}B_{1,1}
\end{eqnarray} and
\begin{eqnarray}
D_{2,1}&=&\frac{(2m+1)A_{0,0}-1}{(2m+1)A_{0,0}+3}B_{2,1}\nonumber\\ &&+\frac{6B_{0,0}B_{2,1}}{[(2m+1)A_{0,0}+3][(2m+1)A_{0,0}+4]}\nonumber\\
&&-\frac{8[(2m+1)A_{0,0}+1]B_{0,0}B_{1,1}^{2}}{[(2m+1)A_{0,0}+3]^{3}[(2m+1)A_{0,0}+4]}
\\
C_{2,1}&=&\frac{8[(2m+1)A_{0,0}+1]B_{1,1}^{2}}{[(2m+1)A_{0,0}+3]^{3}[(2m+1)A_{0,0}+4]}\nonumber\\
       &&+\frac{(2m+1)A_{0,0}-2}{(2m+1)A_{0,0}+4}A_{2,1}
\end{eqnarray}
with
\begin{eqnarray}
&&R_{0;m}(A_{0,0})=(2m+1)A_{0,0}+1,\\
&&R_{1;m}(A_{0,0},B_{0,0},B_{1,1})=-\frac{4B_{0,0}B_{1,1}}{(2m+1)A_{0,0}+3}
\\
&&R_{2;m}(A_{0,0},B_{0,0},B_{1,1},B_{2,1},A_{2,1})\nonumber\\
&=&[-\frac{4B_{0,0}B_{2,1}}{(2m+1)A_{0,0}+3}+AB_{1,1}^{2}+BA_{2,1}]
\end{eqnarray}
where
\begin{eqnarray}
A=\frac{8B_{0,0}^{2}-8[(2m+1)A_{0,0}-1][(2m+1)A_{0,0}+3]}{[(2m+1)A_{0,0}+3]^{3}[(2m+1)A_{0,0}+4]}\nonumber\\
B=\frac{6B_{0,0}^{2}-2[(2m+1)A_{0,0}-1][(2m+1)A_{0,0}+3]}{[(2m+1)A_{0,0}+3][(2m+1)A_{0,0}+4]}
\end{eqnarray}
For the general proof of $n$, one uses the induction methods to
proceed. The above result shows that the formula
(\ref{shape-invariance v-pm }) is true for $N=0,\ 1,2$. Suppose it
is true for $N\le n-1$, we need to prove it is also true for $N=n$.
First, one simplifies the expressions of
$V^{\pm}_n(A_{i,j},B_{i,j},\theta)$. With the help of
\begin{eqnarray}
&&W^2=\sum_{n=0}^{\infty}\beta^n\sum_{k=0}^{n}W_k(A_{i,j},B_{n,j},\theta)W_{n-k}(A_{i,j},B_{n,j},\theta)
\end{eqnarray}
we have the formulae for $V_n^{\pm}$ in the case $n\ge 1$ as
following
\begin{eqnarray}
V^{-}_n(A_{i,j},B_{n,j},\theta)&=&P_{n}^{-}+\bar{P}_{n}\\
V^{+}_n(A_{i,j},B_{n,j},\theta)&=&P_{n}^{+}+\bar{P}_{n}
\end{eqnarray}
There are three parts in the above equations and we simplify them
separately.
\begin{eqnarray}
P_{n}^{-}&=&2W_0W_n(A_{n,j},B_{n,j},\theta)-W'_n(A_{i,j},B_{n,j},\theta)\\
&=&\cos\theta\sum_{p=1}^{[\frac{n+1}{2}]}P^{-}_{n,p}(A_{i,j},B_{n,j})\sin^{2p-2}\theta\nonumber\\
&&
+\sum_{p=1}^{[\frac{n+2}{2}]}Q^{-}_{n,p}(A_{i,j},B_{n,j})\sin^{2p-2}\theta
\end{eqnarray}
where
\begin{eqnarray}
&&P^-_{n,p}(A_{n,j},B_{n,j})=-B_{0,0}A_{n,p}a_{n,p}\nonumber\\
&&+\bigg[(1-2p)-(2m+1)A_{0,0}\bigg]B_{n,p}b_{n,p}
\end{eqnarray}
\begin{eqnarray}
&&Q^-_{n,p}(A_{n,j},B_{n,j})=-B_{0,0}B_{n,p}b_{n,p}\nonumber\\
&& \ \ \ + \bigg[(1-2p)-(2m+1)A_{0,0}\bigg]A_{n,p}a_{n,p}
\nonumber\\
&& \ \ \  +\bigg[(2p-2)+(2m+1)A_{0,0}\bigg]A_{n,p-1}a_{n,p-1}
\end{eqnarray}
\begin{eqnarray}
P_{n}^{+}&=&2W_0W_n(A_{n,j},B_{n,j},\theta)+W'_n(A_{i,j},B_{n,j},\theta)\\
&=&\cos\theta\sum_{p=1}^{[\frac{n+1}{2}]}P^{+}_{n,p}(A_{i,j},B_{n,j})\sin^{2p-2}\theta\nonumber\\
&&
+\sum_{p=1}^{[\frac{n+2}{2}]}Q^{+}_{n,p}(A_{i,j},B_{n,j})\sin^{2p-2}\theta
\end{eqnarray}
where
\begin{eqnarray}
&&P^+_{n,p}(A_{n,j},B_{n,j})=-B_{0,0}A_{n,p}a_{n,p}
\nonumber\\
&&+ \bigg[(2p-1)-(2m+1)A_{0,0}\bigg]B_{n,p}b_{n,p}
\end{eqnarray}
\begin{eqnarray}
&&Q^+_{n,p}(A_{n,j},B_{n,j})=-B_{0,0}B_{n,p}b_{n,p}
\nonumber\\
&&+
\bigg[-(2m+1)A_{0,0}+(2p-1)\bigg]A_{n,p}a_{n,p}\nonumber\\
&& +\bigg[(2-2p)+(2m+1)A_{0,0}\bigg]A_{n,p-1}a_{n,p-1}
\end{eqnarray} Notice the fact
\begin{eqnarray}
 &&P^-_{n,p}(A_{n,j},B_{n,j})=P^+_{n,p}(A_{n,j},B_{n,j})\nonumber\\
 && =
Q^-_{n,p}(A_{n,j},B_{n,j})
 =Q^+_{n,p}(A_{n,j},B_{n,j})=0,\nonumber\\
 && \ \ \ \ \ \ \ \   \ p<1\ or\  p> [\frac{n+1}2] \label{P+-np,Q-np=0 condition}.
\end{eqnarray}
It is easy to obtain \cite{ground of frac12}
\begin{eqnarray}
\bar{P}_{n}&=&\sum_{k=1}^{n-1}\bigg[W_k(A_{i,j},B_{n,j},\theta)W_{n-k}(A_{i,j},B_{n,j},\theta)\bigg]\nonumber\\
&=&\cos\theta\sum_{p=1}^{[\frac{n+1}{2}]}G_{n,p}(A_{i,j},B_{i,j})\sin^{2p-2}\theta\nonumber\\
&&+\sum_{p=1}^{[\frac{n+2}{2}]}H_{n,p}(A_{i,j},B_{i,j})\sin^{2p-2}\theta
\end{eqnarray}
where
\begin{eqnarray}
G_{n,p}(A_{i,j},B_{i,j})&=&\sum_{k=1}^{n-1}\sum_{j=1}^{p-1}
\bigg[\bar{b}_{k,p-j}\bar{a}_{n-k,j}+\bar{a}_{k,p-j}\bar{b}_{n-k,j}\bigg]\label{b-p2}\\
H_{n,p}(A_{i,j},B_{i,j})&=&\sum_{k=1}^{n-1}\sum_{j=1}^{p-1}
\bigg[\bar{b}_{k,p-j}\bar{b}_{n-k,j}+\bar{a}_{k,p-j}\bar{a}_{n-k,j}\nonumber\\
&&-\bar{a}_{k,p-1-j}\bar{a}_{n-k,j}\bigg]\label{b-p3}
\end{eqnarray}
where
\begin{eqnarray}
\bar{a}_{i,j}=A_{i,j}a_{i,j},\  \bar{b}_{i,j}=B_{i,j}b_{i,j}.
\end{eqnarray}
When $n$ is even,
\begin{eqnarray}
  G_{n,p}(A_{i,j},B_{i,j})&=&0,\  \ p<2\  or\  p> \frac n2;\  \nonumber\\
   H_{n,p}(A_{i,j},B_{i,j})&=&0,\  \ p<2\ or\  p> \frac n2+1,
  \end{eqnarray}
and when $n$ is odd,
\begin{eqnarray}G_{n,p}(A_{i,j},B_{i,j})&=&H_{n,p}(A_{i,j},B_{i,j})=0,\label{G-np,H-np=0 condition}.
\end{eqnarray}
 Thus,
\begin{eqnarray}
&&V^-_n(A_{i,j},B_{n,j})\nonumber\\ &=&\cos\theta\sum_{k=2}^{[\frac{n+1}{2}]}[G_{n,p}(A_{i,j},B_{i,j})+P^-_{n,p}(A_{n,j},B_{n,j})]sin^{2p-2}\theta\nonumber\\
&&\ \ +\sum_{k=2}^{[\frac{n+2}{2}]}[H_{n,p}(A_{i,j},B_{i,j})+Q^-_{n,p}(A_{n,j},B_{n,j})]sin^{2p-2}\theta\nonumber\\
&&\ \ +P^-_{n,1}(A_{n,j},B_{n,j})+Q^-_{n,1}(A_{n,j},B_{n,j})
\label{vn-}\\
&&V^+_n(A_{i,j},B_{n,j})\nonumber\\ &=&\cos\theta\sum_{k=2}^{[\frac{n+1}{2}]}[G_{n,p}(A_{i,j},B_{i,j})+P^+_{n,p}(A_{n,j},B_{n,j})]sin^{2p-2}\theta\nonumber\\
&&\ \ +\sum_{k=2}^{[\frac{n+2}{2}]}[H_{n,p}(A_{i,j},B_{i,j})+Q^+_{n,p}(A_{n,j},B_{n,j})]sin^{2p-2}\theta\nonumber\\
&&\ \ +P^+_{n,1}(A_{n,j},B_{n,j})+Q^+_{n,1}(A_{n,j},B_{n,j})
\label{vn+}
\end{eqnarray}
one could rewrite $V_n^+(A_{i,j})$ as
\begin{eqnarray}
&&V^+_n(A_{i,j},B_{i,j})=V^-_n(C_{i,j},D_{i,j})+R_{n;m}(A_{i,j},B_{i,j})\nonumber\\
&=&\cos\theta\sum_{k=2}^{[\frac{n+1}{2}]}[G_{n,p}(C_{i,j},D_{i,j})+P^-_{n,p}(C_{n,j},D_{n,j})]sin^{2p-2}\theta\nonumber\\
&&+\sum_{k=2}^{[\frac{n+2}{2}]}[H_{n,p}(C_{i,j},D_{i,j})+Q^-_{n,p}(C_{n,j},D_{n,j})]sin^{2p-2}\theta\nonumber\\
&&+P^-_{n,1}(C_{n,j},D_{n,j})+Q^-_{n,1}(C_{n,j},D_{n,j})+R_{n;m}(A_{i,j},B_{i,j})
\end{eqnarray}
\begin{eqnarray}
R_{n;m}(A_{i,j},B_{i,j})&=&P^+_{n,1}(A_{n,j},B_{n,j})+Q^+_{n,1}(A_{n,j},B_{n,j})\nonumber\\
&&-P^-_{n,1}(C_{n,j},D_{n,j})-Q^-_{n,1}(C_{n,j},D_{n,j})
\end{eqnarray}
where
\begin{eqnarray}
P^-_{n,1}(C_{n,j},D_{n,j})=-[(2m+1)C_{0,0}+1]C_{n,1}b_{n,1}\\
Q^-_{n,1}(C_{n,j},D_{n,j})=-D_{0,0}C_{n,1}b_{n,1}
\end{eqnarray}
and
\begin{eqnarray}
P^+_{n,1}(A_{n,j},B_{n,j})=-[(2m+1)A_{0,0}-1]B_{n,1}b_{n,1}\\
Q^+_{n,1}(A_{n,j},B_{n,j})=-B_{0,0}B_{n,1}b_{n,1}
\end{eqnarray}
by the use of
\begin{eqnarray}
b_{n,0}=0\qquad a_{n,0}=0\qquad n=1,2,\dots
\end{eqnarray}
 In order to maintain the shape-invariance
property for the nth term, the following equations must be satisfied
\begin{eqnarray}
&&G_{n,p}(C_{i,j},D_{i,j})+P^-_{n,p}(C_{n,j},D_{n,j})\nonumber\\
&&=G_{n,p}(A_{i,j},B_{i,j})+P^+_{n,p}(A_{n,j},B_{n,j})\nonumber\\
&&H_{n,p}(C_{i,j},D_{i,j})+Q^-_{n,p}(C_{n,j},D_{n,j})\nonumber\\
&&=H_{n,p}(A_{i,j},B_{i,j})+Q^+_{n,p}(A_{n,j},B_{n,j})\nonumber\\
&& \ \ \ \ \ \ \ \ p=2,3,\ldots. [\frac{n+2}2]\nonumber
\end{eqnarray}
Define
\begin{eqnarray}
 &&G_{n,p}(A_{i,j},B_{i,j})-G_{n,p}(C_{i,j},D_{i,j})+P^+_{n,p}(A_{n,j},B_{n,j})\nonumber\\
 && \ \ \equiv U_{n,p},\\
 &&H_{n,p}(A_{i,j},B_{i,j})-H_{n,p}(C_{i,j},D_{i,j})+Q^+_{n,p}(A_{n,j},B_{n,j}) \nonumber\\ && \ \ \equiv
 \check{U}_{n,p},
\end{eqnarray}
then we have
\begin{eqnarray}
P^-_{n,p}(C_{n,j},D_{n,j})&=& -\alpha_{p}D_{n,p}b_{n,p}-D_{0,0}C_{n,p}a_{n,p}\nonumber\\ &=&U_{n,p}\nonumber\\
Q^-_{n,p}(C_{n,j},D_{n,j})&=&
-\alpha_{p}C_{n,p}a_{n,p}-D_{0,0}D_{n,p}b_{n,p}\nonumber\\ && +
(\alpha_{p}-1)C_{n,p-1}a_{n,p-1}\nonumber\\ &=& \check{U}_{n,p},
\end{eqnarray}
where $\alpha_{p}=\bigg[(2m+1)C_{0,0}+(2p-1)\bigg]$.

 In the above equations, the quantities $C_{i,j},i<n,\
j<[\frac{i}2]$  and $D_{i,j},i<n,\ j<[\frac{i+1}2]$ are known as the
functions of the variables $A_{i,j},B_{i,j},\ i<n,\ j<n$. The only
unknown quantities are the $n$ quantities $C_{n,p},\ p=1,2,\ldots,\
[\frac{n}2] $, $D_{n,p},\ p=1,2,\ldots,\ [\frac{n+1}2] $. Therefore,

From the above equations, one obtains
\begin{eqnarray}
D_{n,p}&=&\frac{D_{0,0}a_{n,p}}{\alpha_{p}b_{n,p}}C_{n,p}-\frac{U_{n,p}}{\alpha_{p}b_{n,p}}\label{d-np}\\
C_{n,p-1}&=&
\frac{\bigg(\alpha_{p}+\frac{D^2_{0,0}}{\alpha_{p}}\bigg)a_{n,p}}{(\alpha_{p}-1)a_{n,p-1}}C_{n,p}+
\frac{\check{U}_{n,p}-\frac{D_{0,0}}{\alpha_{p}}U_{n,p}}
{(\alpha_{p}-1)a_{n,p-1}},\label{c-np}\\
&& \ \ \ \ \ \ \ \ \ p=2,3,\ldots, [\frac{n+2}2]
\end{eqnarray} with $D_{n,[\frac{n+1}2]+1}=C_{n,[\frac{n}2]+1}=0$.
Here we give some notes: (1) when $n$ is odd, one could obtain first
$D_{n,[\frac{n+1}2]}$ from Eq.(\ref{d-np}) and $C_{n,[\frac{n}2]}$
from Eq.(\ref{c-np}) under the condition $p=[\frac{n+1}2],\
C_{n,[\frac{n+1}2]}=0$. Then, it is easy to calculate subsequently.
(2) when $n$ is even, one needs to calculate $C_{n,\frac{n}2}$ from
Eq.(\ref{c-np}) under the condition $p=[\frac{n+1}2],\
C_{n,[\frac{n+1}2]}=0$. Then, it is easy to calculate subsequently.

 Then, the  excited eigen-values $E_{l;m}$and
eigenfunctions $\Psi_l$ is obtained by the recurrence relation :
\begin{eqnarray}
&&  E_{l;m}^-=E_{0;m}+\sum_{k=1}^{l}R(a_{k},b_{k}),\\&&
E_{0;m}=m(m+1)-\frac{3}{4}+\sum_{n=1}^{\infty}E_{0,n;m}\beta^n
\\
&&
R(a_{k},b_{k})=R_{0;m}+\sum_{n=1}^{\infty}\beta^nR_{n;m}(a_{k},b_{k}),\\
&&a_1=(A_{i,j},B_{i,j}),\ a_2=(C_{i,j},D_{i,j}),\ldots, \\ &&
\Psi_0\propto
\exp\bigg[-\int_{\theta_0}^{\theta}W(A_{n,j},B_{n,j},\theta)d\theta\bigg],
\\ && {\cal A}^{\dagger}=-\frac d{d\theta}+W(A_{n,j},B_{n,j},\theta)\\
&& \Psi_n^-={\cal A}^{\dagger}
\,(A_{n,j},B_{n,j},\theta)\Psi_{n-1}^-(C_{n,j},D_{n,j},\theta),\\ &&
\  \ \ \ \ \ \ \ \ \ \  \ n=1,2,3,\dots
\end{eqnarray}
In conclusion, we have proved the shape-invariance properties for
the spin-weighted equations in the case of $s=\frac12$ and obtain
the recurrence relations for them. By these results we can get the
exited eigenvalue and eigenfunction. Similar process could also
extends to the case $s=1,2,\frac32$ for the spin-weighted functions.

\section*{Acknowledgements}
 This work is supported by the National Natural Science Foundation of China (Grant Nos.10875018,10773002)


\begin{thebibliography}{99}
\bibitem{Teu1} S.A.Teukolsky. Rotating Black Holes: Separable Wave
Equations for Gravitational and Electromagnetic Perturbations.
Phys.Rev.Lett, 1972, 29: 1114.

\bibitem{Teu2} S.A.Teukolsky. Perturbations of a Rotating Black Hole. I.
Fundamental Equations for Gravitational, Electromagnetic, and
Neutrino-Field Perturbations. Astrophys. J, 1973, 185: 635.

\bibitem{Flammer3} Flammer C. Spheroidal wave functions. Stanford,
CA£ºStandford Univiersity Press,1956.

\bibitem{Flammer4}
Tian G H. Integral Equations for the Spin-Weighted Spheroidal Wave
Functions. Chin.Phys.Lett, 2005, 22: 3013

\bibitem{5}
Marc Casals and Adrain C.ottewill, High frequency asymptotics for
the spin-weighted spheroidal equation.Phys. Rev. D , 2005,71:064025,
see also references cited there.

\bibitem{Stratton4} Tian G H  New Investigation of Spheroidal Wave Functions,
Chin. Phys. Lett, 2010, 27: 030308.

\bibitem{Li5}Tian G H and Zhong S Q. Ground State Eigenfunction of
Spheroidal Wave Functions, Chin. Phys. Lett, 2010, 27: 040305.

\bibitem{Tian6}Tian G H and Zhong S Q. Investigation of the recurrence
relations for the spheroidal wave functions, Arxiv, 2009, 0906.4687
V3

\bibitem{Tian7}Tian G H and Zhong S Q.Ground State Eigenfunction of
Spheroidal Wave Functions, Chin. Phys. Lett, 2010, 27: 100306

\bibitem{Tian8}Tang W L and Tian G H. Solve the spheroidal wave equation
with small c by SUSYQM method, accept by Chin. Phys. B, 2010
preprint
\bibitem{Infeld9} Zhou J, Tian G H and Tang W L. The spin-weighted
spheroidal wave functions in the case of $s=\frac12$, J.
Math.Phys,2010 (submitted).
\bibitem{Cooper11}Li K, Sun Y, Tian G H and Tang W L. The spin-weighted
spheroidal wave functions in the case of $s=2$, accept by Chin. Sci.
G,2010, preprint (In Chinese)
\bibitem{Tian12} Cooper F, Khare A, Sukhatme U 1995
\bibitem{Berti13}Gradsbteyn I.S., Ryzbik L.M. Table of integrals, series,
and products. 6th ed. Singapore:Elsevierpte.Ltd, 2000


\end{thebibliography}
\end{document}